\documentstyle[preprint,aps,epsf]{revtex}
\begin{document}
\draft
\preprint{TIT/HEP-260/COSMO-45}
\title{Topology Changes by Quantum Tunneling in Four Dimensions}
\author{Ding Shuxue\footnote{e-mail:sding@phys.titech.ac.jp}, Yasushige 
Maeda\footnote{e-mail:y-maeda@hoffman.cc.sophia.ac.jp} and Masaru Siino\footnote{e-mail:msiino@phys.titech.ac.jp, 
JSPS fellow}\\
\it ${}^{*\ddagger}$Department of Physics, Tokyo Institute of Technology\\ 
Meguroku, Tokyo 152, Japan\\
${}^{\dagger}$Department of Mathematics, Faculty of Science and Technology,\\
Sophia University, 7-1 Kioi-cho, Chiyodaku, Tokyo 102, Japan}

\maketitle
\begin{abstract}
We investigate topology-changing processes in 4-dimensional quantum gravity with 
a negative cosmological constant. By playing the ``gluing-polytope game" 
in hyperbolic geometry, we explicitly construct an instanton-like 
solution without singularity. Because of cusps, this solution is non-compact but has a 
finite volume. Then we evaluate a topology change amplitude in the 
WKB approximation in terms of the volume of this solution.
\end{abstract}

Topology change may occur in quantum 
gravity though it would not happen in physically restricted classical 
spacetimes\cite{GST}. In 3-dimensional spacetime with a negative 
cosmological constant, Fujiwara, Higuchi, 
Hosoya, Mishima and one of the present authors(M.~S.)\cite{FHHMS} demonstrated that the 
topology change can occur due to the quantum tunneling effect by constructing 
the explicit examples of the solutions. According to Gibbons and Hartle\cite{GH}, the 
quantum tunneling spacetime is semi-classically approximated by
a Riemannian manifold with totally geodesic boundaries. In 
Ref.\cite{FHHMS} such manifolds are constructed from regular truncated 
polyhedra embedded in a hyperbolic 3-space. In this paper we extend 
their procedure to a 4-dimensional spacetime and construct a 
solution of the Einstein equation. Using 4-dimensional regular truncated 
polytope embedded in a hyperbolic 4-space\cite{TH}, we construct a 
4-dimensional spacetime solution corresponding to the topology change by a quantum tunneling effect.

Gibbons and Hartle\cite{GH} proposed that the tunneling process is 
described by a Riemannian manifold which has the boundaries $\Sigma_i$ 
and $\Sigma_f$ in the WKB approximation. The topology change is characterized 
by the difference of topologies of the initial spatial 
hypersurface $\Sigma_i$ and the final spatial hypersurface $\Sigma_f$. For 
quantum tunneling in the semi-classical picture, these spatial hypersurfaces are required to have vanishing extrinsic curvatures.
We call these boundary hypersurfaces with vanishing extrinsic curvature as 
totally geodesic boundaries\cite{GH}.

\begin{figure}[htb]
	\centerline{\epsfxsize=8.5cm \epsfbox{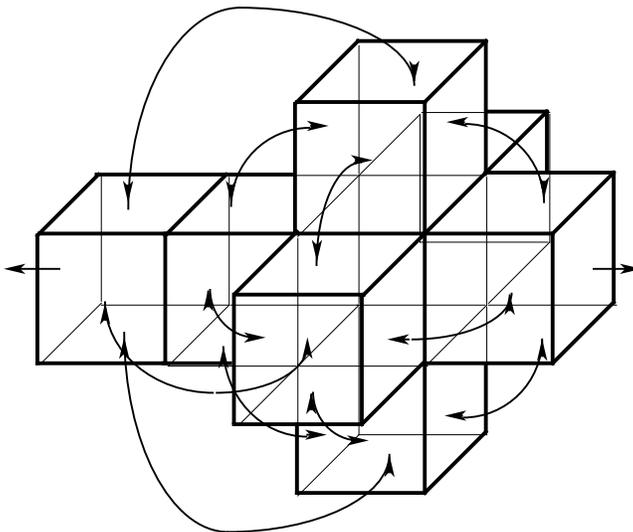}}
	\caption{The development of an 8-cell. Gluing the faces of the hexahedra along the arrows in four 
	dimensions, we get the 8-cell of a 4-dimensional polytope.}
	\protect\label{fig:unf}
\end{figure}

 When we suppose that the spacetime  
is homogeneous and the Weyl tensor vanishes everywhere, the Riemannian manifold becomes locally 
isometric to any one of the cases of $S^4$ (4-sphere), $R^4$ (4-plane) or $H^4$ 
(4-hyperboloid). In Ref.\cite{GH} it is also 
stated that if we require that the $\Sigma_i$ and $\Sigma_f$ are 
disconnected, the spacetime should at some points violate an energy 
condition, which demands
\begin{equation}
	R_{\mu \nu}V^{\mu}V^{\nu}>0
	\label{}
\end{equation}
for all vector $V^{\mu}$. Therefore we can exclude $S^4$ from our 
considerations of topology changing manifold because the 
curvature is positive. Since the variety 
of hyperbolic manifolds (Riemannian manifolds locally isometric to $H^4$) 
is very rich, we shall consider the vacuum case  
with a negative cosmological constant. Then the following question arises;
\begin{quote}
Can we construct a hyperbolic 4-manifold with totally geodesic boundaries 
$\Sigma_i$ and $\Sigma_f$ which have different topologies?
\end{quote}
In this case it is noted that from the Gauss-Codazzi equation,
the vanishing extrinsic curvature makes $\Sigma_i$ and $\Sigma_f$ also 
have a hyperbolic structure (locally isometric to $H^3$).
Any 4-manifold with hyperbolic structure is the quotient manifold 
of a hyperbolic 4-space with a discrete subgroup of its isometry group $SO(4,1)$. The 
fundamental region of this quotient 4-manifolds is a 4-polytope embedded 
into $H^4$. The boundaries of the fundamental region are identified with 
each other and the fundamental region forms a 4-manifold. Following the procedure of 3-dimensional 
case\cite{FHHMS}, we determine the fundamental region and the 
identifications of 
its boundary faces in the hyperbolic geometry. Our steps 
to construct the 4-dimensional solution are following: 
\begin{description}
\item[Step (1):] Decide 4-polytopes which we shall use. 
\item[Step (2):] Give them a hyperbolic structure and truncate its vertices. 
\item[Step (3):] Find identifications of faces on 3-boundaries made by the truncation so 
as to form 3-manifolds. 
\item[Step (4):]
Find the identifications of polyhedra bounding the 4-polytope, which induces 
the identifications found in the step (3) on the 3-boundaries.
\end{description}
 The step (3) is 
trivial in 3-dimensional case. While all 
2-manifolds with hyperbolic structure are classified as Riemann surfaces, 
we know only a small number of hyperbolic 3-manifolds. This step makes our trial 
non-trivial. In the step (1), we use a more or less systematic way to decide 
4-polytopes which will be explained in a separate
 paper\cite{NEXT}. Here we decide to use twelve 
 8-cell's\cite{COX} which are 
 4-polytopes bounded by eight congruent hexahedra. The development of an 
 8-cell on 3-space is shown in Fig.\ref{fig:unf}. Gluing faces in 
 four dimensions
 according to the arrows depicted in Fig.\ref{fig:unf} we get a 4-dimensional polytope 
 surrounded by these eight hexahedra, which has sixteen vertices.

To give a hyperbolic structure to these twelve 8-cell's, we embed them into a
hyperbolic 4-space $H^4$. Here we shall use an n-dimensional projective 
model of a hyperbolic n-space\cite{TH}, where a totally geodesic 
hypersurface is a hyperplane in the sense of Euclidean geometry. The 
n-projective model is a model on an open n-disk
\begin{equation}
	D^n=\{x^i\in R^n\vert x^i x_i<1\},
	\label{}
\end{equation}
in which a metric is
\begin{equation}
	ds^2 = {1 \over 1-r^2}\left({dr^2 \over 1-r^2} +r^2 
	d\Omega_{n-1}^2\right).
	\label{}
\end{equation}
When $r$ goes to $1$, one approaches to a sphere at infinity $\partial D^n$. 
This metric gives a constant sectional curvature $-1$. In this model all 
 gluing procedures of hyperplanes are executed by isometries $SO(n,1)$ of the 
 hyperbolic n-space. 
 
\begin{figure}[htb]
	\centerline{\epsfxsize=8.5cm \epsfbox{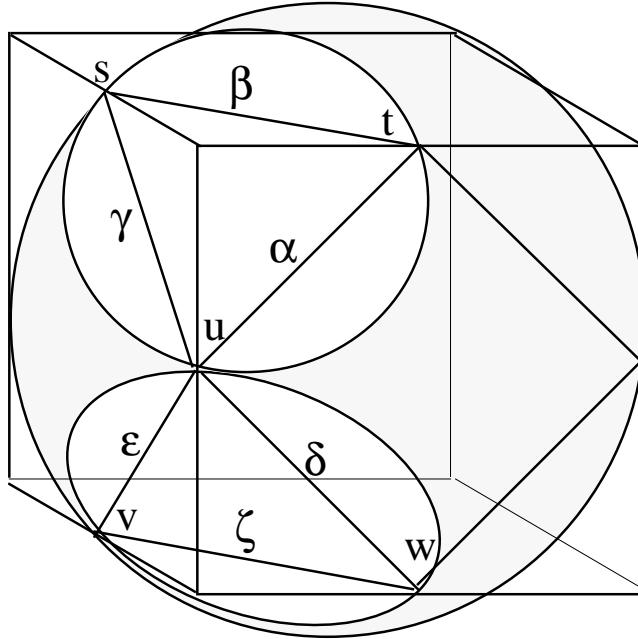}}
	\caption{The shaded sphere is a sphere at infinity of a 3-projective model. Each edge of the 
	hexahedra is tangent to the sphere at $s,t,u...$. The sphere is cut by 
	planes through $s,t,u...$. Along these planes we 
	truncate the vertices of the hexahedron.}
	\protect\label{fig:cus2}
\end{figure}

\begin{figure}[htb]
	\centerline{\epsfxsize=8.5cm \epsfbox{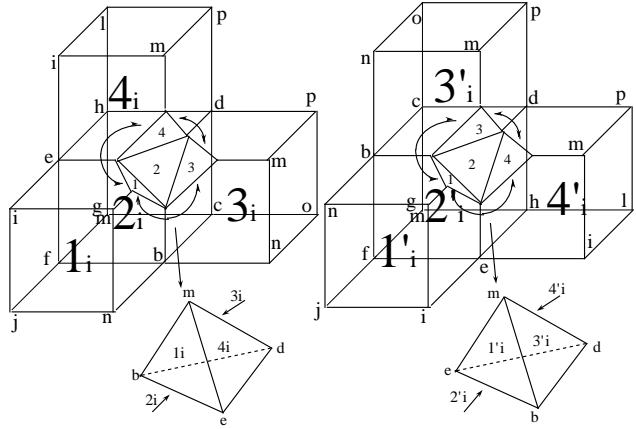}}
	\caption{Four hexahedra meet at a vertex. When we truncate the vertex, 
	a tetrahedron appears.  We label the faces of a 
tetrahedra by the cells which the face 
belongs to, and the vertices of a tetrahedra by the nearest neighbore vertices of the 8-cell's.}
	\protect\label{fig:trt}
\end{figure}

 The size of the embedded 8-cell is determined so that its dicellular 
 angles (the angle between two adjacent hexahedra in 4-dimensions) 
 becomes $\pi/3$ (the angle decreases as the size increases in the 
 hyperbolic geometry). 
 It is noted that in this size every vertex is out of the sphere at infinity $\partial 
 D^4$ and edges of the 8-cell are tangent to the sphere. Then, each 
 hexahedron of the 8-cell is embedded into an induced 3-projective model 
 (sub-model of the 4-projective model) as shown 
 in Fig.\ref{fig:cus2}. As it shows, every vertex is 
out of the sphere at infinity $\partial D^3$ and all edges are 
 tangent to the sphere.
In the same way as the 3-dimensional case \cite{FHHMS}, we truncate each 
vertex of the 8-cell. Let us pay attention to the four hexahedra having a 
vertex in common. A remarkable property of the hyperbolic space guarantees 
the existence of a unique 3-hyperplane which is perpendicular to all of 
the four hexahedra. We cut the sixteen vertices of the 8-cell along these 3-hyperplanes 
to get a regular truncated 8-cell embedded completely in 
the 4-projective model. These truncations of 8-cell induce truncations of 
each hexahedron bounding the 8-cell. The resultant hexahedron is shown in 
Fig.\ref{fig:cus2}. On the section of the open 3-disk a triangle appears
and its three vertices are on 
the sphere at infinity $\partial D^3$. It is noted that the triangles 
share vertices with neighboring triangles and every original edge of the 
hexahedron is completely truncated off by two adjacent truncations. 
Since a single 8-cell has sixteen 
vertices, the truncated polytope contains sixteen boundary components made by 
the truncation. Because four hexahedra share one vertex in an 8-cell and the triangles 
made by the truncation form a tetrahedron, we see that the boundary is a tetrahedron whose vertices are 
on the sphere at infinity (see Fig.\ref{fig:trt}). The dihedral angle of 
the tetrahedron is $\pi/3$ due to the regular truncation. An easy calculation 
tells us that such tetrahedra have a finite volume though they are 
non-compact.

\begin{figure}[htb]
	\centerline{\epsfxsize=8.5cm \epsfbox{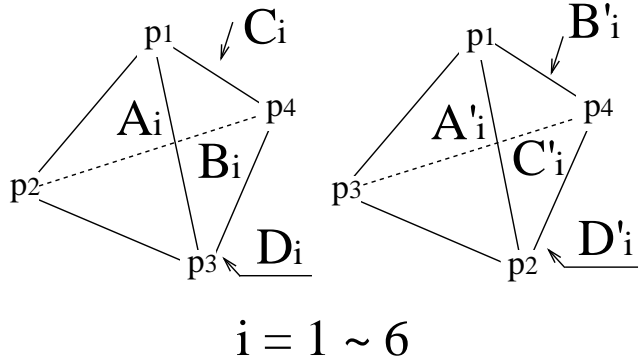}}
	\caption{We consider six left-handed tetrahedra and six right-handed 
	tetrahedra. Twelve tetrahedra form ${\cal M}_b$.}
	\protect\label{fig:fekc}
\end{figure}

For the step (3), we consider a 3-manifold on the boundary, which should be 
constructed from several tetrahedra with a dihedral angle $\pi/3$ and their vertices on the sphere 
at infinity $\partial D^3$ but have a finite volume.
In this paper, we construct a suitable manifold ${\cal M}_b$ from such twelve 
tetrahedra, which is a 
non-compact but complete smooth 3-manifold with a finite volume. The faces 
and vertices of the twelve tetrahedra are labeled as depicted in Fig.\ref{fig:fekc}. 
It should be noticed that the six primed tetrahedra are the mirror reflections 
of the other unprimed six tetrahedra. We call these primed tetrahedra as 
right-handed and the unprimed as left-handed.
The following pairs of the faces of right-handed tetrahedra and the faces 
of the left-handed tetrahedra are glued so that each labeled vertex 
matches. 
\begin{equation}
\begin{array}{cccc}
	A_1-A'_1 & B_1-B'_3 & C_1-C'_2 & D_1-D'_4  \\
	A_2-A'_2 & B_2-B'_1 & C_2-C'_4 & D_2-D'_3  \\
	A_3-A'_3 & B_3-B'_2 & C_3-C'_5 & D_3-D'_6  \\
	A_4-A'_4 & B_4-B'_6 & C_4-C'_1 & D_4-D'_5  \\
	A_5-A'_5 & B_5-B'_4 & C_5-C'_6 & D_5-D'_1  \\
	A_6-A'_6 & B_6-B'_5 & C_6-C'_3 & D_6-D'_2
\end{array}
\label{eqn:gl1}	
\end{equation}
For instance, $A_1$ is matched with $A'_1$. All the 
vertices $p_1, p_2, p_3$ of $A_1$  are identified with the 
vertices $p_1, p_2, p_3$ of $A'_1$, respectively.
 Since the dihedral angle is $\pi/3$, a 
neighborhood of each point on edge is a 3-ball (This can be checked 
by exhaustion in the way executed in Ref.\cite{FHHMS}). 

On the other hand, the vertices of the tetrahedra are on the sphere at 
infinity $\partial D^3$. Hence these vertices form four 
cusps\cite{cusp} corresponding to the vertices $p_1, p_2, p_3, 
p_4$. Then ${\cal M}_b$ is non-compact. We, however, are not 
disappointed by this non-compactness since the manifold has a finite volume and is complete and 
smooth. There 
is no singularity. If there were no cusps, we should also check the 
consistency of identifications additionally on each vertex about solid angles. This 
consistency check makes the 
searching of solutions much involved. In fact, we know much more cusped 
3-manifolds than compact 3-manifolds. Hence, admitting the cusp in our 
manifold we get the following simplest example of topology change 
solutions.

\begin{figure}[htb]
	\centerline{\epsfxsize=8.5cm \epsfbox{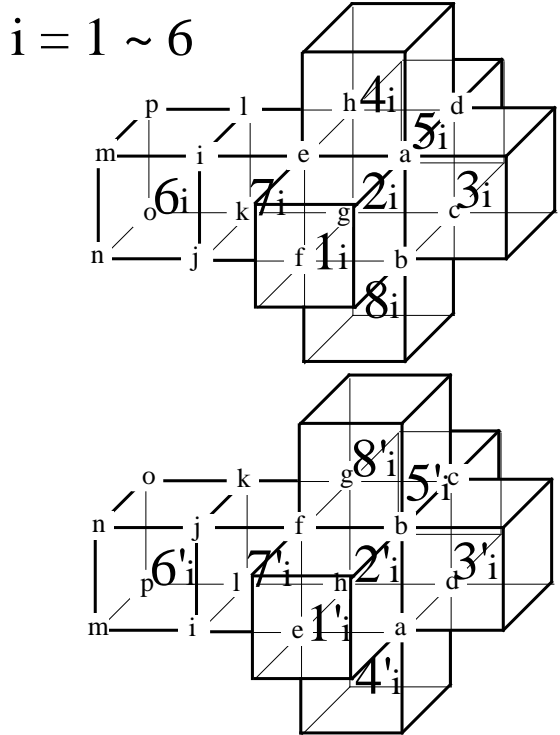}}
	\caption{There are two types of 8-cell's. Upper ones are with 
	left-handed tetrahedra. 
	Lower ones (with prime) are with right-handed tetrahedra. The corresponding cells (for 
	example, $1_1$ and $1'_1$, $4_1$ and $4_3'$ ...) are identified.}
	\protect\label{fig:const}
\end{figure}

In the step (4) we expect that there are appropriate identifications 
among cells (hexahedra) 
of twelve 8-cell's, which induce identifications on the $16\times 12$ tetrahedra 
given by the truncation of the vertices of the twelve 8-cell's (possessing sixteen 
vertices) so as to form sixteen ${\cal 
M}_b$'s. We find such identifications of 
the cells, as follows. We divide the twelve 8-cell's into six 
left-handed ones and six right-handed ones with prime 
(Fig.\ref{fig:const})
following the twelve tetrahedra composing ${\cal M}_b$ (see 
Fig.\ref{fig:fekc}).
Each vertex and cell (hexahedra) of the 8-cell's are labeled as shown in
Fig.\ref{fig:const}. One identifies 
the pairs of the cells (hexahedra) of the left-handed 8-cell's 
and that of the right-handed 8-cell's so that the labeled vertices 
$(a\sim p)$ match as shown below. 
The pairs are determined following the gluing of the tetrahedra composing 
${\cal M}_b$ (\ref{eqn:gl1}).
\begin{equation}
\begin{array}{cccc}
	1_1-1'_1 & 4_1-4'_3 & 3_1-3'_2 & 2_1-2'_4  \\
	1_2-1'_2 & 4_2-4'_1 & 3_2-3'_4 & 2_2-2'_3  \\
	1_3-1'_3 & 4_3-4'_2 & 3_3-3'_5 & 2_3-2'_6  \\
	1_4-1'_4 & 4_4-4'_6 & 3_4-3'_1 & 2_4-2'_5  \\
	1_5-1'_5 & 4_5-4'_4 & 3_5-3'_6 & 2_5-2'_1  \\
	1_6-1'_6 & 4_6-4'_5 & 3_6-3'_3 & 2_6-2'_2  \\
	5_1-5'_1 & 8_1-8'_3 & 7_1-7'_2 & 6_1-6'_4  \\
	5_2-5'_2 & 8_2-8'_1 & 7_2-7'_4 & 6_2-6'_3  \\
	5_3-5'_3 & 8_3-8'_2 & 7_3-7'_5 & 6_3-6'_6  \\
	5_4-5'_4 & 8_4-8'_6 & 7_4-7'_1 & 6_4-6'_5  \\
	5_5-5'_5 & 8_5-8'_4 & 7_5-7'_6 & 6_5-6'_1  \\
	5_6-5'_6 & 8_6-8'_5 & 7_6-7'_3 & 6_6-6'_2
\end{array}
\label{eqn:gl2}
\end{equation}
 Of course these
identifications are orientation preserving isometry transformation 
because of reflection 
symmetry between the primed 8-cell's and the others.
The resultant space is orientable.
By these identifications, twelve vertices with the same name (for example, 
twelve vertices named $(a)$ in each 8-cell in Fig.\ref{fig:const})
are all identified. 
Then corresponding twelve tetrahedra made by the truncation of these 
twelve vertices with the same name 
 are glued. The gluing of the faces of the tetrahedra is determined by the 
gluing of the cells. For example, labeling the faces of the 
tetrahedron named $(a)$ by the index of the cell which the face 
belongs to, and the vertices of the tetrahedron $(a)$ by the 
index of the nearest neighbore vertices of the 8-cell's, 
these twelve tetrahedra are equivalent to the tetrahedra composing ${\cal 
M}_b$ (see Fig.\ref{fig:trt} and Fig.\ref{fig:fekc}). Comparing the two gluings
(\ref{eqn:gl1}) and (\ref{eqn:gl2}), we 
find that the twelve tetrahedra $(a)$ form ${\cal M}_b$ since the 
both gluings are done so that the labeled vertices of the tetrahedra match.

It is easy to check that 
the tetrahedra made by the truncations of the vertices with the other names 
$(b)\sim (p)$ 
also form ${\cal M}_b$. Then, the 4-space constructed from the  8-cell has sixteen 
totally geodesic boundaries ${\cal M}_b$'s, since a regular truncation guarantees that the boundary given 
by the truncation is a totally geodesic smooth manifold\cite{FHHMS}.

To check that this 4-space is 
complete smooth 4-manifold, we consider the neighborhood of faces, edges and 
vertices. In 4-dimensions, when we turn around each face completely the 
total angle has to be $2\pi$ by consistency. 
On the boundary 3-hypersurface, however, this amounts to checking the 
$2\pi$ turn around the edges ($\alpha, \beta, \gamma ...$ in 
Fig.\ref{fig:cus2}) of the boundary. This consistency is 
guaranteed because the boundary is an already checked manifold ${\cal M}_b$.
The remaining vertex after the regular 
truncation ($s, t, u ...$ in Fig.\ref{fig:cus2}) causes no problem since they form
 4-cusps at infinity. The edges are located only on 
the boundaries which form 3-manifold after the gluing ($\alpha, \beta, \gamma ...$ in 
Fig.\ref{fig:cus2}). As mentioned above, this does not bring any trouble.
Hence this space is a hyperbolic complete smooth 4-manifold with totally 
geodesic 3-boundaries 
possessing a hyperbolic structure. The boundaries are sixteen ${\cal 
M}_b$'s. 

\begin{figure}[htb]
	\centerline{\epsfxsize=8.5cm \epsfbox{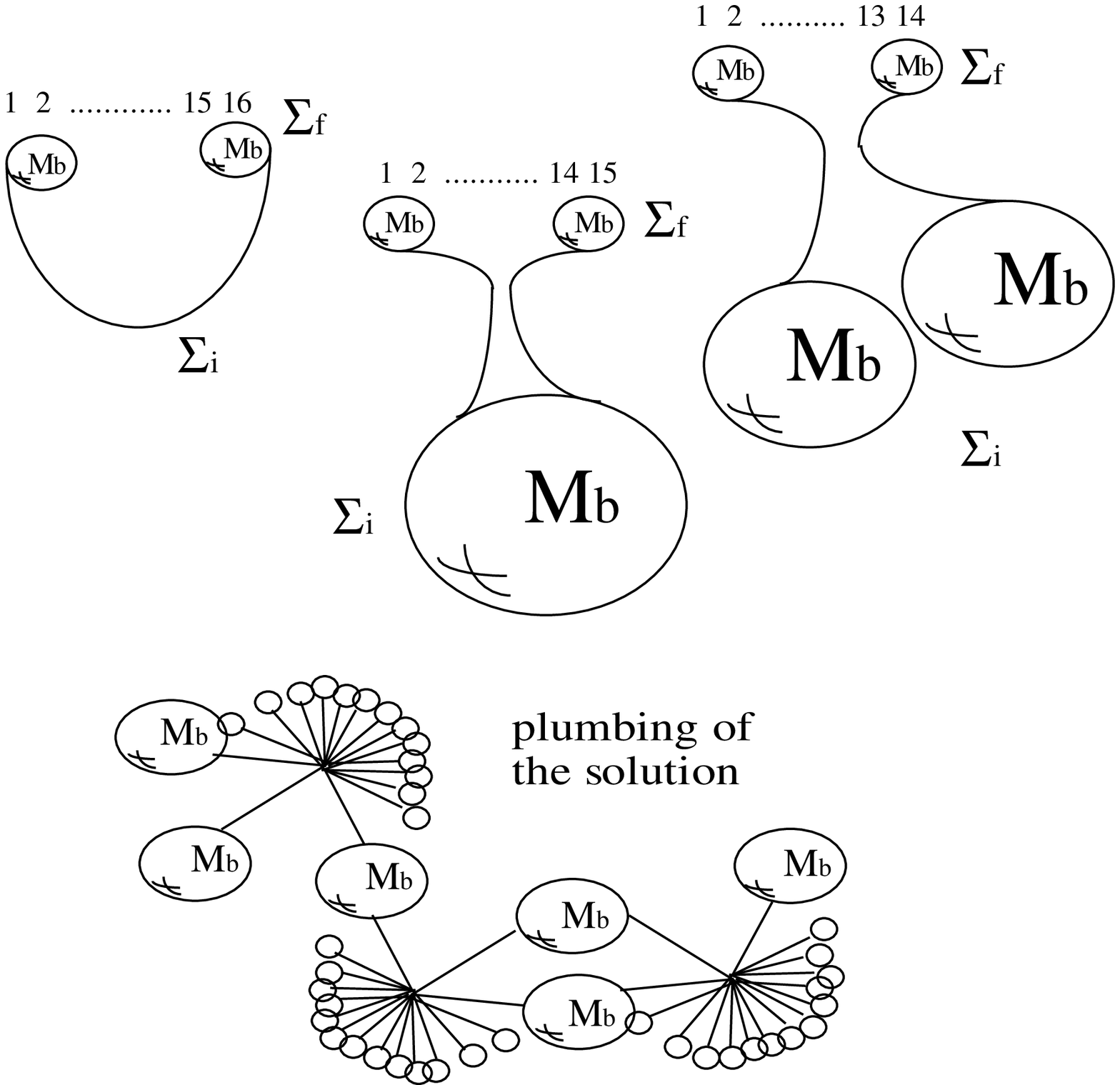}}
	\caption{The Riemannian manifold with eight boundaries is regarded as 
	the topology change solution `from nothing to sixteen ${\cal M}_b$'s',
	 `from one ${\cal M}_b$ to fifteen ${\cal M}_b$'s' or `from two ${\cal M}_b$'s to fourteen ${\cal M}_b$'s', and so on. Furthermore, by plumbing of the solution we 
get various types of topology change solutions.}
	\protect\label{fig:plu}
\end{figure}

From Gibbons and Hartle\cite{GH} and Fujiwara, Higuchi, Hosoya, Mishima and  
one of the present authors (M.~S.)\cite{FHHMS} we can see that this manifold is 
regarded as an instanton causing topology change by quantum tunneling, for example, 
`from nothing to sixteen ${\cal M}_b$'s', `from one ${\cal M}_b$ to 
fifteen ${\cal M}_b$'s' or `from two ${\cal M}_b$'s to fourteen ${\cal 
M}_b$'s', and so on (see Fig.\ref{fig:plu}). It is also worthy 
of notice 
that by plumbing them we can get infinite series of topology change solutions 
as examplified in  Fig.\ref{fig:plu}.
The topology change shown in this paper is 
the first explicit example of a topology change in four dimensions by quantum 
tunneling effect, which cannot be reduced to a lower-dimensional 
spacetime. Brill\cite{DB} investigated the topology changing 4-spacetime which 
is essentially a two dimensional topology change.
 Though the manifold does not satisfy the `no boundary' 
boundary condition rigorously, we assume that the amplitude can be formally described 
by the Hawking's Riemannian path integral as 
\begin{equation}
	T(h_i, h_f)=\sum_{M_R}\int{\cal D}g \exp(-S_E[g]),
	\label{eqn:path}
\end{equation}
where $h_i$ and $h_f$ are the 3-dimensional metrics on the initial 
spatial hypersurface $\Sigma_i$ and the final spatial hypersurface 
$\Sigma_f$, respectively. $S_E$ is the Euclidean action,
\begin{equation}
	S_E=-{1 \over 16\pi G}\int(R-2\Lambda)\sqrt{g}d^4x.
	\label{}
\end{equation}
The path integral is over smooth 4-metric $g$ on the Riemannian manifold 
$M_R$ which has appropriate boundaries $\Sigma_i$ and $\Sigma_f$ by 
assumption. For the present case, it turns out that $M_R$ is a 4-manifold 
bounded by sixteen ${\cal M}_b$'s. Then we can use the obtained solution to evaluate the path 
integral (\ref{eqn:path}) in the WKB approximation. Since our solution 
has a constant negative 
curvature $R=4\Lambda$, the classical action $\bar{S}_E$ is given by
\begin{equation}
\bar{S}_E={1 \over 8\pi G}{V \over \vert \Lambda \vert},
	\label{}
\end{equation}
where $V$ is a numerical value representing the volume of $M_R$ in the 
case of $\Lambda=-1$.
 Though our manifold has cusps, the volume is 
finite. The calculation of the volume will be shown in forthcoming paper\cite{NEXT}. 

People might be disturbed by the existence of the cusps. However, we can 
argue that it 
physically causes no trouble since the cusps are at infinity and we cannot 
see. We see only the pattern of spatial periodicity\cite{FS}.
If we 
take the position that the cusp is not allowed, the construction becomes 
more difficult. In this case we may use a computational calculation\cite{NEXT}.

We would like to thank Professor A. Hosoya and Professor S. Kojima for helpful discussions.
One of the authors (M. S.) thanks the Japan Society for the 
Promotion of Science for financial support. This work was supported in 
part 
by the Japanese Grant-in-Aid for Scientific Research Fund of the Ministry 
of 
Education, Science and Culture.


\end{document}